\documentclass[twocolumn,aps,pra]{revtex4}

\begin{document}

\title{Comment on ''Surface-impedance approach solves problems with the thermal
Casimir force between real metals''}
\author{V. B. Svetovoy }
\email[]{V.B.Svetovoy@el.utwente.nl}
\thanks{On leave from Yaroslavl University, Yaroslavl, Russia}
\affiliation{Transducers Science and Technology Group, EWI,
University of Twente, P.O. 217, 7500 AE Enschede, The Netherlands}
\date{\today}

\begin{abstract}
In a recent paper, Geyer, Klimchitskaya, and Mostepanenko [Phys.
Rev. A \textbf{67}, 062102 (2003)] proposed the final solution of
the problem of temperature correction to the Casimir force between
real metals. The basic idea was that one cannot use the dielectric
permittivity in the frequency region where a real current may
arise leading to Joule heating of the metal. Instead, the surface
impedance approach is proposed as a solution of all
contradictions. The purpose of this comment is to show that (i)
the main idea contradicts to the fluctuation dissipation theorem,
(ii) the proposed method to calculate the force gives wrong value
of the temperature correction since the contribution of low
frequency fluctuations is calculated with the impedance which is
not applicable at low frequencies. In the impedance approach the
right result for the reflection coefficients in the $n=0$ term of
the Lifshitz formula is given.
\end{abstract}


\maketitle

The problem of temperature dependence of the Casimir force between
real metals gave rise to a vivid discussion in the literature
\cite{BS00,SL,BGKM00,GLR00,KM01,TL02,BAH02,BKM02,SL03,HBAM03,GKM03}
(see \cite{GKM03} for more extensive reference list). Direct
application of the Lifshitz formula \cite{LP9} to calculate the
force between two metallic plates showed \cite{BS00} that the
transverse electric mode does not contribute to the force in the
zero frequency limit. The reflection coefficients in this limit
was found to be independent on the metal properties:

\begin{equation}
r_{\Vert }^{2}\left( 0,q\right) =1,\;r_{\bot }^{2}\left(
0,q\right) =0, \label{RM}
\end{equation}

\noindent where $\mathbf{q}$ is the wave vector along the plates
and the indexes $\Vert $ and $\bot $ stand for parallel
(transverse magnetic) and perpendicular (transverse electric)
polarizations. The problem is that for the ideal metal both
polarizations contribute equally, $r_{\Vert }^{2}=r_{\bot
}^{2}=1$, and there is no continuous transition between real and
ideal metals. Difference in the reflection coefficients provides
difference in the temperature corrections to the Casimir force.
For the ideal metal this correction is negligible at small
separations between bodies but for real metals with coefficients
(\ref{RM}) it becomes large and disagrees with the torsion
pendulum experiment \cite{Lam97}.

Different research groups took part in the discussion of the
problem but after a few years still there is no a satisfactory
solution. It can be a signal that we do not understand something
important in good old electrodynamics and it makes the problem
even more interesting.

In the commented paper \cite{GKM03} B. Geyer, G. Klimchtskaya, and V.
Mostepanenko claim that they have found the final solution of the problem.
Before going into details let us present the structure of the Lifshitz
formula that will be used in the discussion. The Casimir free energy per
unit area for two metallic semispaces separated by the distance $a$ is

\begin{eqnarray}
\mathcal{F}=\frac{kT}{2\pi }\sum\limits_{n=0}^{\infty }{}^{\prime
}\int\limits_{\zeta _{n}/c}^{\infty }\left[ \ln \left( 1-r_{\Vert
}^{2}\left( \zeta _{n},q\right) \exp (-2qa)\right) +\right.
\label{Lif} \nonumber\\
\left. \left( r_{\Vert }\rightarrow r_{\bot }\right) \right] qdq,
\end{eqnarray}

\noindent where $\zeta _{n}=2\pi kTn/\hbar $ are the Matsubara
frequencies. Only the reflection coefficients depend on the
material properties, which are described by the dielectric
function $\varepsilon \left( i\zeta _{n}\right) $ at imaginary
frequencies.

The driving idea of Ref. \cite{GKM03} is that the electromagnetic
fluctuations resulting in Eq.(\ref{Lif}) cannot be used with the
Drude dielectric function

\begin{equation}
\varepsilon \left( \omega \right) =1-\frac{\omega _{p}^{2}}{\omega \left(
\omega +i\omega _{\tau }\right) },  \label{Drude}
\end{equation}

\noindent where $\omega _{p}$ and $\omega _{\tau }$ are the plasma and
relaxation frequencies, because non-zero imaginary part of $\varepsilon
\left( \omega \right) $ will give rise to a real current inside of metal and
will produce heating of the metal.

In this connection we would like to stress that any fluctuating
physical value in the system which is in thermal equilibrium is
connected with the dissipation via the fluctuation dissipation
theorem. The energy for fluctuation is taken from the heat bath
and returned back with the dissipation of fluctuation. No heat is
produced in the process. The fluctuations of electromagnetic field
are not an exception to this fundamental theorem of statistical
physics. The simplest example is the Johnson-Nyquist noise. Quite
real fluctuating in time current can be measured in a wire without
application any external force. Spectral density of this current
is proportional to the temperature and wire conductivity. Lifshitz
has used the theory of electromagnetic fluctuations \cite{Lif56}
to get his famous formula for the Casimir force. The fluctuation
dissipation theorem lies in the heart of this theory. In this
approach the force is defined by the material dielectric function
at imaginary frequencies $\varepsilon \left( i\zeta \right) $.
This function cannot be measured directly but can
be expressed via the imaginary part of $\varepsilon \left( \omega \right) =$ $%
\varepsilon ^{\prime }\left( \omega \right) +i$ $\varepsilon ^{\prime \prime
}\left( \omega \right) $ using the dispersion relation
\begin{equation}
\varepsilon \left( i\zeta \right) =1+\frac{2}{\pi }\int\limits_{0}^{\infty }%
\frac{\omega \varepsilon ^{\prime \prime }\left( \omega \right) }{\omega
^{2}+\zeta ^{2}}d\omega .  \label{DispIm}
\end{equation}

\noindent In this sense one can say that the force itself is defined by the
dissipation \cite{LP9}. It is true even for the plasma model, for which $%
\omega _{\tau }$ in (\ref{Drude}) is going to zero. We cannot
completely
neglect $\omega _{\tau }$ since the dispersion relations between $%
\varepsilon ^{\prime }\left( \omega \right) $ and $\varepsilon ^{\prime
\prime }\left( \omega \right) $ will be broken. It is obvious that $%
\varepsilon ^{\prime \prime }\left( \omega \right) $ is also important in
relation (\ref{DispIm}). Therefore, we have to keep $\omega _{\tau }$
arbitrary small but finite. It is known effect for the fluctuations in a
lossless medium \cite{LP9}.

We can conclude that the main idea of the commented paper
contradicts to the physical nature of fluctuations. However, the
surface impedance approach the authors consider as an alternative
has its right to exist. It has been already discussed in a number
of papers \cite{MT85,BKR02,SL03} and it is definitely preferable
at least in the range of anomalous skin effect \cite{SL03}. In
this approach the optical properties of a metal are
described with the surface impedance $Z\left( \omega \right) $ instead of $%
\varepsilon \left( \omega \right) $. The boundary conditions connect the
tangential components of electric $\mathbf{E}_{t}$ and magnetic $\mathbf{H}%
_{t}$ fields on the surface. For this reason the electromagnetic
problem is much simpler to solve since there is no need to solve
it inside of metal. Of course, in this case the Casimir force is
also connected with the dissipation of surface current that
authors seem did not notice. This current is defined as

\begin{equation}
\mathbf{j}_{S}=\left[ \mathbf{H}_{t}\,\mathbf{n}\right] ,
\end{equation}

\noindent where $\mathbf{n}$ is the normal unit vector directed
inside of metal. Via the impedance boundary condition this current
is connected with the surface electric field

\begin{equation}
\mathbf{j}_{S}=\frac{\mathbf{E}_{t}}{Z\left( \omega \right) }\,.
\end{equation}

\noindent Real part of $Z\left( \omega \right) $ is responsible for
dissipation.

To calculate the Casimir force, the authors are using the Lifshitz formula (%
\ref{Lif}) with the reflection coefficients expressed via the impedance at
imaginary frequencies $Z\left( i\zeta \right) $ (see Eq.(41) \cite{GKM03}%
). Unfortunately, the way they perform the calculations also
cannot be considered as satisfactory. The function $Z\left( \omega
\right) $ can be represented in different explicit forms depending
on the frequency range. The authors separate the domains of normal
skin effect, anomalous skin effect, and domain of infrared optics.
To make calculations they use these pieces of the same function in
the following peculiar way: ''... at each separation distance
between the plates one should, first, determine the characteristic
frequency $\omega _{c}$ and, second, fix the proper impedance
function. Thereafter the chosen impedance function can be used at
all frequencies when performing the integration in Eq.(48). At
zero temperature this prescription is optional. At $T\neq 0$,
however, it takes on great significance.'' The characteristic
frequency $\omega _{c}$ mentioned here is defined as $\omega
_{c}=c/2a$, where $a$ is the separation between plates, Eq.(48)
gives the Casimir energy (free energy (\ref{Lif}) in this text at
$T\rightarrow 0$).

At this point we have to make a comment. The surface impedance
knows nothing about the problem in which we are going to use it.
It is one function pieces of which we know in explicit analytic
form for different frequency domains. It depends only on the
material properties. At any circumstances this function cannot
change in dependence on distance between plates. One has to use it
as a whole for the force calculation.

It is easy to understand what will happen if we proceed as the authors of
the commented paper recommend. Suppose $\omega _{c}$ corresponds to the
infrared optics domain where

\begin{equation}
Z\left( \omega \right) =-i\frac{\omega }{\sqrt{\omega _{p}^{2}-\omega ^{2}}}.
\label{IR}
\end{equation}

\noindent The authors continue this function to small frequencies, where Eq. (%
\ref{IR}) does not work any more. This procedure is equivalent to a definite
prescription for the $n=0$ term in the Lifshitz formula (\ref{Lif}). Really,
the reflection coefficients at $\omega =i0$ with the impedance given by (\ref
{IR}) are (see Eq.(58) \cite{GKM03})

\begin{equation}
r_{\Vert }^{2}\left( 0,q\right) =1,\;r_{\bot }^{2}\left( 0,q\right) =\left(
\frac{\omega _{p}-cq}{\omega _{p}+cq}\right) ^{2}.  \label{PMRef}
\end{equation}

\noindent Since (\ref{IR}) is the impedance of the plasma model,
the coefficients (\ref{PMRef}) reproduce the prescription of the
plasma model \cite{BGKM00}. Negligible temperature correction at
small separations between bodies is predicted with the
coefficients (\ref{PMRef}) \cite{BGKM00}.

The right way to calculate the Casimir force is straightforward
and quite obvious. The contribution of fluctuations with a
frequency $\omega $ must be taken into account with the impedance
$Z\left( \omega \right) $ which corresponds to the same frequency.
When $\omega _{c}$ is in the infrared optics domain most of the
terms in the Lifshitz formula should be taken with the impedance
(\ref{IR}) but the fluctuations with $\omega \ll \omega _{\tau }$
must be calculated with an appropriate impedance of the normal
skin effect

\begin{equation}
Z\left( \omega \right) =\left( 1-i\right) \sqrt{\frac{\omega }{8\pi \sigma }}%
,  \label{NSE}
\end{equation}

\noindent where $\sigma $ is the material conductivity. In the
Lifshitz formula (\ref{Lif}) always there is at least one term
($n=0$) for which we have to use Eq. (\ref{NSE}). Then instead of
(\ref{PMRef}) for the reflection coefficients at $\omega =i0$ we
will find

\begin{equation}
r_{\Vert }^{2}\left( 0,q\right) =1,\;r_{\bot }^{2}\left( 0,q\right) =1.
\label{IMRef}
\end{equation}

\noindent These coefficients coincide with that for the ideal
metal. Temperature correction in this case will be small but not
negligible \cite{SL}.

The essence of the problem with the temperature correction for
real metals is the value of the coefficient $r_{\bot }^{2}\left(
0,q\right) $. Together with M. Lokhanin the author of this comment
provided a number of physical arguments \cite{SL,SL03} in favor of
(\ref{IMRef}). However, as was emphasized in the conclusion of
Ref. \cite{SL03}, we still unsure that the impedance approach
gives the final solution. At room temperature the normal skin
effect is applicable at low frequencies and it is unclear why the
impedance boundary conditions should be preferable in comparison
with the conditions of continuity of tangential components of
electric and magnetic fields on the surface.

Our last comment is related to the range of anomalous skin effect. A good
guiding principle to deal with the temperature correction was proposed in
Ref. \cite{BKM02}. Any physically reasonable result must obey the Nernst
heat theorem: the entropy must disappear in the zero temperature limit $%
T\rightarrow 0$. At low temperatures and frequencies most of good
metals are in the range of anomalous skin effect. In our paper
\cite{SL03} it was noted for the first time that to check the
Nernst theorem one has to use the impedance of anomalous skin
effect for calculations. We
also demonstrated that the entropy is going to zero in the limit $%
T\rightarrow 0$ only if the reflection coefficients are chosen as
in (\ref {IMRef}). Additionally it was shown that the relative
temperature correction to the free energy is not negligibly small.

The authors of commented paper came to a different conclusion that
the temperature correction in the anomalous skin effect domain is
very small. They connected disagreement with the improper range of
application for the anomalous skin effect in our paper. In reality
the difference comes again
from the way the they calculated the Casimir force. At small separations $%
a=100-500\;nm$ considered in our paper the characteristic
frequency in the gap between plates is in the range of infrared
optics $\omega _{c}=\left( 0.3-1.5\right) \cdot 10^{15}\;rad/s$
and the main contribution in the force is really comes from the
impedance (\ref{IR}). The important terms in the Lifshitz formula
correspond to large $n\sim \hbar \omega _{c}/2\pi kT\gg 1
$. The same is not true for the temperature correction. Much smaller $%
n<\hbar \Omega /2\pi kT$ give the main contribution to the
temperature correction \cite{SL03}, where $\Omega =v_{F}\omega
_{p}/c$ is the characteristic frequency of anomalous skin effect.
This condition is equivalent to the Matsubara frequencies $\zeta
_{n}<\Omega $ which are in the anomalous skin effect domain with
the impedance

\begin{equation}
Z\left( i\zeta \right) =\left( \frac{v}{c}\frac{\zeta ^{2}}{\omega _{p}^{2}}%
\right) ^{1/3},  \label{ASE}
\end{equation}

\noindent where $v$ is of the order of the Fermi velocity $v_{F}$.
Instead of using (\ref{ASE}) the authors \cite{GKM03} continued
Eq. (\ref{IR}) to
the frequencies  $\zeta _{n}<\Omega $ where it is not applicable. Since (\ref{IR}%
) corresponds to the plasma model and it is known that the
temperature correction in this model is very small, they got their
result. However, this result has no relation with reality.

In conclusion, we see that the original claim \cite{GKM03} for the
final solution of the problem with the temperature correction
cannot be supported. The inconsistent way to calculate the Casimir
force did not allow to derive the right result for the temperature
correction. Actually the authors introduced a specific
prescription from the very beginning.

However, the commented paper so as the previous works
\cite{BKR02,SL03} draw our attention to the surface impedance
approach. At low temperatures and low frequencies it is the only
way to describe a metal. At higher temperatures and low
frequencies it seems that both descriptions with the dielectric
function and surface impedance are equally right but they give
completely different results for the temperature correction. The
impedance approach is in good correspondence with the ideal metal,
but the usual boundary conditions do not
give smooth transition to the ideal metal. As was pointed out in Ref. \cite{HBAM03}%
, it is a natural conclusion since in the static limit magnetic
field penetrates freely into a real metal but does not penetrate
into the ideal metal. We still do not know the solution of this
problem.

\begin{acknowledgments}
The author is grateful to the Transducer Science and Technology
Group, Twente University for hospitality. This work was supported
by the Dutch Technology Foundation.
\end{acknowledgments}


\begin{thebibliography}{99}
\bibitem{BS00} M. Bostr\"{o}m and B. E. Sernelius, Phys. Rev. Lett.
\textbf{84}, 4757 (2000).

\bibitem{SL} V. B. Svetovoy and M. V. Lokhanin, Mod. Phys. Lett. A
\textbf{15},1013 (2000); 1437 (2000); Phys. Lett. A \textbf{280},
177 (2001).

\bibitem{BGKM00} M. Bordag, B. Geyer, G.L. Klimchitskaya, and V.
M. Mostepanenko, Phys. Rev. Lett. \textbf{85}, 503 (2000).

\bibitem{GLR00} C. Genet, A. Lambrecht, and S. Reynaud, Phys. Rev.
A \textbf{62}, 012110 (2000).

\bibitem{KM01} G. L. Klimchitskaya and V. M. Mostepanenko, Phys.
Rev. A \textbf{63}, 062108 (2001).

\bibitem{TL02} J. R. Torgenson ans S. K. Lamoreaux,
quant-ph/0208042.

\bibitem{BAH02} I. Brevik, J. B. Aarseth, and J. S. H\o ye, Phys.
Rev. E \textbf{66}, 026119 (2002).

\bibitem{BKM02} V. B. Bezerra, G. L. Klimchitskaya, and V. M.
Mostepanenko, Phys. Rev. A \textbf{65}, 052113 (2002).

\bibitem{SL03} V. B. Svetovoy and M. V. Lokhanin, Phys. Rev. A
\textbf{67}, 022113 (2003).

\bibitem{HBAM03} J. S. H\o ye, I. Brevik, J. B. Aarseth, and K. A. Milton, Phys.
Rev. E \textbf{67}, 056116 (2003).

\bibitem{GKM03} B. Geyer, G.L. Klimchitskaya, and V.
M. Mostepanenko, Phys. Rev. A \textbf{67}, 062102 (2003).

\bibitem{LP9} E. M. Lifshitz and L. P. Pitaevskii, \emph{Statistical
Physics}, Part II (Pergamon Press, Oxford, 1980).

\bibitem{Lam97} S. K. Lamoreaux, Phys. Rev. Lett. \textbf{78}, 5
(1997).

\bibitem{Lif56} E. M. Lifshitz, Zh. Eksp. Teor. Fiz. \textbf{29},
94 (1956) [Sov. Phys. JETP \textbf{2}, 73 (1956)].

\bibitem{MT85} V. M. Mostepanenko and N. N. Trunov, Yad. Fiz.
\textbf{42}, 1297 (1985) [Sov. J. Nucl. Phys. \textbf{42}, 818
(1985)].

\bibitem{BKR02} V. B. Bezerra, G. L. Klimchitskaya, and C. Romero, Phys. Rev. A \textbf{65}, 012111 (2002).

\end{thebibliography}
\end{document}